# DISCUSSION OF: TREELETS—AN ADAPTIVE MULTI-SCALE BASIS FOR SPARSE UNORDERED DATA

By Catherine Tuglus and Mark J. van der Laan

*University of California, Berkeley*

We would like to congratulate Lee, Nadler and Wasserman on their contribution to clustering and data reduction methods for high $p$ and low $n$ situations. A composite of clustering and traditional principal components analysis, treelets is an innovative method for multi-resolution analysis of unordered data. It is an improvement over traditional PCA and an important contribution to clustering methodology. Their paper presents theory and supporting applications addressing the two main goals of the treelet method: (1) Uncover the underlying structure of the data and (2) Data reduction prior to statistical learning methods. We will organize our discussion into two main parts to address their methodology in terms of each of these two goals. We will present and discuss treelets in terms of a clustering algorithm and an improvement over traditional PCA. We will also discuss the applicability of treelets to more general data, in particular, the application of treelets to microarray data.

**1. Uncover the underlying structure of the data.** In order to determine the underlying structure of a given data set, the statistician will often employ various clustering algorithms, or projection-based methods such as principal components analysis in an effort to tease apart the data which is often highly correlated and very noisy. The authors, Lee, Nadler and Wasserman, propose a new method targeted at detecting the multi-resolution internal structure of the data. In wavelet-fashion, the results are presented on multiple scales, providing detail only when necessary. However, unlike wavelet-analysis, their technique is applicable to unordered data. Though presented initially as an extension of wavelets, treelets are built upon a hierarchical clustering framework and can be illustrated as such.

As outlined in the overview van der Laan, Pollard and Bryan (2003), clustering methods are described by three major components: the distance measure, the grouping criteria, and the algorithm. The authors in this paper present treelets in terms of a correlation distance matrix, while we have argued for algorithms which allow arbitrary distance metrics since different applications can require different uses of the notion of proximity. Though they elude that other distance measures can be applied, all theory and simulation is presented and proven using a covariance or correlation measure of similarity. When alternate distance measures are used the benefit of using this method over other clustering methods seems questionable, and the final interpretation of the multi-resolution basis is unclear.

When the underlying structure of the data does not reflect a sparse diagonal correlation matrix, using more adaptable clustering methods such as Hierarchical Partitioning and Collapsing Hybrid (HOPACH) [Pollard and van der Laan (2005), van der Laan and Pollard (2003)] would be more appropriate and seem to provide more flexibility and more interpretable results. HOPACH takes as input an arbitrary distance or dissimilarity matrix, combines top-down and agglomerative clustering into a hybrid algorithm, allows for data adaptively deciding on the number of children cluster in each node, orders the clusters in each layer of the hierarchical tree based on the distance so that neighboring clusters are close to each other w.r.t. the specified dissimilarity, and it allows the use of a data adaptive as well as visual criteria (including output of bootstrap) to decide on the depth and number of clusters in the tree.

The treelet algorithm is a binary agglomerative hierarchical clustering algorithm. In terms of a hierarchical graph only, the two most correlated nodes are combined at a given step. For an $n$ by $p$ data matrix, there are total $p-1$ layers for a graph combined to completion. The binary combination allows for the multi-resolution interpretability of the resulting basis. At each node a principal components analysis is applied to the pair of variables. The node is then represented by the two components, the first component becoming a "sum" variable, and the second the "difference" variable. Since only the sum variable is allowed to combine in higher levels of the graph, the difference variable remains behind as a residual measure of the combination. Each treelet, comprised of one node (sum variable) and its associated difference variables can be represented by a orthonormal basis.

The treelet method is applicable given any agglomerative hierarchical algorithm. However, the graph is solely built on the similarity between two variables. This does not take advantage of all information present in the data. Clustering algorithms have advanced beyond simple similarity measures and use informative measures such as the Mean Silhouette [Kaufman and Rousseeuw (1990)], the Median Silhouette, or the Split Mean/Median Silhouette [van der Laan, Pollard and Bryan (2003)]. Each of these grouping criteria reflects how similar variables are in relation to how dissimilar they are from others.

The authors do present a measure to determine the optional height of the tree, a normalized energy score reflecting the percent variance explained on a given basis conditional on the number of variables chosen to represent the treelet—the best $K$-dimensional basis. According to the authors, the best height and dimension K can be chosen using cross-validation - though the exact method of cross validation is not presented clearly in terms of choosing $K$. If the goal is to use treelets for the purpose of prediction, then this is easily defined, but it becomes unclear what is meant otherwise.

In terms of a clustering algorithm, we applaud the authors for having a well defined goal: estimation of the true correlation matrix. Generally cluster analysis, though built from localized structure, does not identify that as its far-reaching goal leaving consistency theory nonexistent. We would like to point out that in terms of clustering, a particular consistency theory for the estimation of the mean and covariance matrix based on Bernstein's inequality, as well as the sensitivity and reproducibility of the estimate based on bootstrap resampling, was presented in van der Laan and Bryan (2001) and subsequent articles.

Beyond a clustering interpretation, treelets can also be viewed as an improved robust version of PCA. Traditional PCA is a global method, highly sensitive to noise in the data. Treelets focus on detecting localized structure and by performing binary data-driven rotations, are much more robust to noise. The authors show the improved finite sample properties of treelets over traditional PCA, and we believe this is a fundamental contribution to the field. Treelets will be able to perform well in many practical settings, while PCA will often rely on too large sample sizes. Treelets also incorporate hierarchical clustering giving the method a wavelet-like property, preserving detailed structure in only the necessary region, unlike PCA which splits the data into orthogonal projections, each with a linear basis relating to the entire data set.

In terms of detecting the underlying structure of data given a sparse correlation matrix, treelets are a great contribution providing a new summary metric for binary clustering algorithms, and providing a localized PCA. In application, however, the method is potentially limited to only data where the underlying correlation structure is assumed to be sparse, such as many image and spatial analyses. Given a more complex correlation structure, which is often seen in biological data such as microarray data, treelets do not necessarily perform better than clustering or standard PCA. The improvement in convergence rate over PCA is contingent on the sparsity of the correlation matrix.

**2. Data reduction.** In terms of data reduction, treelets are a data-driven method which provides a more concise representation of a data matrix with sparse correlation. Reducing the dimension of the initial data set before applying a learning algorithm can improve the accuracy of the predictor. In the spirit of the super-learning approach [van der Laan, Polley and Hubbard (2007)], involving an aggressive approach for data adaptively selecting among a continuum of different strategies for construction of a prediction, for the purposes of dimension reduction in prediction, we recommend in practice that the height of the tree (L) and the dimension of the basis (K) should be chosen with respect to the cross-validated risk of the prediction in all applications. The authors elude to this.

The practical application of treelets as a dimension reduction technique for high-dimensional microarray data is unclear. Microarray data is generally not sparsely correlated with a nice diagonal block structure. In fact, the correlation structure is often very complex and noisy. Though the treelets may provide a set of summary measures for the data set, the benefit of using these summary measures over those obtained using a traditional PCA for this type of data is not demonstrated. Also, we note that though they present the benefits of using their method as data reduction prior to prediction in Sections 5.1 and 5.3, in the case of the Glob DNA microarray data in Section 5.3 the authors chose to reduce the data prior to the application of treelets using univariate regression. They restrict their data to the 1000 most "significant" genes. The reasons for this initial reduction are not stated, nor are the reasons for the arbitrary cut-off of 1000.

Often the truncation of a data set using a $p$-value cut-off is used to improve computational speed or improve accuracy. Regardless of the reasoning, the use of simple linear regression may not achieve an accurate ranking of "significant" genes. Univariate regression is notorious for detecting false positive genes. Constraining the data to the more "significant" genes may decrease the noise of the data, but it will not decrease the complexity of the correlation structure. We argue the use of targeted variable importance using targeted Maximum Likelihood or comparable double robust locally efficient estimation method would provide a more accurate ranking of the potentially causal genes [Bembom et al. (2007), Tuglus and van der Laan (2008)] than univariate regression. We also argue that if the initial reduction was completed to improve accuracy for the sake of prediction, the cut-off should be chosen with respect to the overall prediction performance. The Golub data, though commonly used to demonstrate prediction methods, is also commonly easy to obtain accurate results. The improvement accuracy of the treelet method over others is difficult to see when in general methods seem to perform so well.

**3. Final comments.** In general we believe treelets to be a great contribution to the field. With respect to clustering methodology, it provides a framework which actively searches for the correct underlying correlation structure. Its improvement over PCA when the correlation matrix is believed to be sparse is also impressive. Given the appropriate data and application, treelets will be a very useful and practical tool for statistical analysis.

Division of Biostatistics
University of California—Berkeley
Berkeley, California 94720
USA
E-mail: ctuglus@berkeley.edu

Division of Biostatistics
Department of Statistics
University of California—Berkeley
Berkeley, California 94720
USA
E-mail: laan@berkeley.edu